\documentclass[pre,showpacs,preprint]{revtex4}

\usepackage{amsmath}

\usepackage{graphicx}
\usepackage{dcolumn}
\usepackage{bm}

\begin{document}

\title{Critical behavior of two freely evolving granular gases separated by an adiabatic piston}
\author{J. Javier Brey and Nagi Khalil}
\affiliation{F\'{\i}sica Te\'{o}rica, Universidad de Sevilla,
Apartado de Correos 1065, E-41080, Sevilla, Spain}
\date{\today }

\begin{abstract}
Two granular gases separated by an adiabatic piston and initially in the same macroscopic state are
considered. It is found that a phase transition with an spontaneous symmetry breaking occurs. When the mass
of the piston is increased beyond a critical value, the piston moves to a stationary position different from the middle of the system. The transition is accurately described by a simple kinetic model that takes into
account the velocity fluctuations of the piston.  Interestingly, the final state is not characterized by the equality of the temperatures of the subsystems but by the cooling rates being the same. Some relevant consequences of this feature are discussed.

\end{abstract}


\maketitle

\section{Introduction}
\label{s1}

The so-called adiabatic piston \cite{Ca63} is a typical example showing the relevance of
fluctuations to explain the macroscopic behavior of a system. It consists of an isolated
cylinder containing two gases separated by an adiabatic piston (no internal degrees of
freedom). Usually, the system is initially prepared with the gases in both compartments
at independent equilibrium states and the piston fixed by a clamp at a given position. Then the
clamp is removed and the piston is free to move without friction with the container. One of the interesting
features of this system is that equilibrium thermodynamics cannot predict the final
position of the piston and the states of the gases to both sides of the piston. Nevertheless,
when fluctuations are taken into account, it follows that the piston moves until the system
relaxes to mechanical and thermal equilibrium with equal pressures and temperatures in  both
compartments. In the last years, the problem has attracted a lot of attention \cite{KVyM00,HyR06,CPPyV07},
mainly stimulated by the seminal paper by Lieb \cite{Li99} and the suitability of the model to
investigate fundamental issues in mesoscopic systems. An illuminating review of the adiabatic piston is given in ref. \cite{GyL06}.

Granular gases, modeled as ensembles of particles colliding inelastically, exhibit many similarities with
molecular gases, but also a rather large number of peculiar behaviors
as a consequence of energy dissipation. In particular, they do not have equilibrium states and are
inherently non-equilibrium systems. For this reason, they have been considered as a proving ground for kinetic
theory and non-equilibrium statistical mechanics \cite{Go03,ByP04}. In this context, an adiabatic piston
separating two granular gases appears as a natural system to investigate the effects of the irreversible
dynamics on the interaction between macroscopic systems. Brito {\em et al.} \cite{BRyvB05} considered the case in
which the two inelastic gases are initially prepared in the same macroscopic state, the homogeneous cooling state, and the particles
collide elastically with the piston. They found that the piston eventually collapsed to one of the sides of the container. This
behavior was explained in terms of the instability of the initial state, leading to
an spontaneous left-right symmetry breaking.

Asymmetric inelastic pistons, with both sides made of different materials with different inelasticities, have also been considered
\cite{CByP08,PTyV10}. In these studies, it has been assumed that the piston moves in an infinite bath of elastic particles either with a Gaussian velocity
distribution \cite{CByP08} or with a velocity distribution in which only two discrete opposite velocities are possible \cite{PTyV10}. Of course,
the two kind of systems, elastic piston and inelastic gases versus  infinite baths of elastic gases colliding inelastically
with the piston, are rather different. In the former case, the system is continuously cooling while in the latter a steady state is reached after some transient.

Here the model of the symmetric elastic piston separating two inelastic gases is revisited. The aim is
to incorporate in the description given in \cite{BRyvB05} the effect of the velocity fluctuations of the piston.
This is the mechanism for which energy is transferred between the left and right hand side gases, leading in
the case of molecular gases to an equilibration of the temperature. It will be shown that some kind of equilibration
also happens in the case of granular gases, although driven by a parameter different from the temperature. The quantity that becomes the same is the cooling rate of the two gases and also of the piston. Quite interestingly, these
conditions lead to the existence of a non-equilibrium phase transition with an spontaneous symmetry breaking, when the mass of the piston exceeds some critical value, which depends on the number of particles and on the inelasticity of the collisions between them. The phase transition is accurately described by a simple kinetic theory that incorporates the effect of the velocity fluctuations of the piston. The results fit into an scenario in which the whole system composed by the gases in both compartments and the piston reaches a cooling state characterized by a unique temperature parameter. The results of ref. \cite{BRyvB05} are recovered in the limit of an infinitely massive piston.

The remaining of the paper is organized as follows. In the next section the system under consideration is specified and the simple theoretical model is formulated. Also the states to be addressed are specified. Evolution equations for the temperature parameters of the granular gases and the piston in those states are written down. Two different kind of states are possible. In one group, the system is symmetric to both sides of the piston, while in the other group the initial left-right symmetry is spontaneously broken. Both states are discussed in Secs. \ref{s3} and \ref{s4}, respectively. A control parameter characterizing whether the state is symmetric or asymmetric is introduced. In Sec.
\ref{s5} the theoretical predictions are compared with molecular dynamics simulations and a good agreement is
observed in all the accesible parameter region. The final section contains some concluding remarks and also a discussion of the possible implications of the reported results for the understanding of the interactions between macroscopic  granular systems.

\section{The model}
\label{s2}
\begin{figure}
\includegraphics[scale=0.5,angle=0]{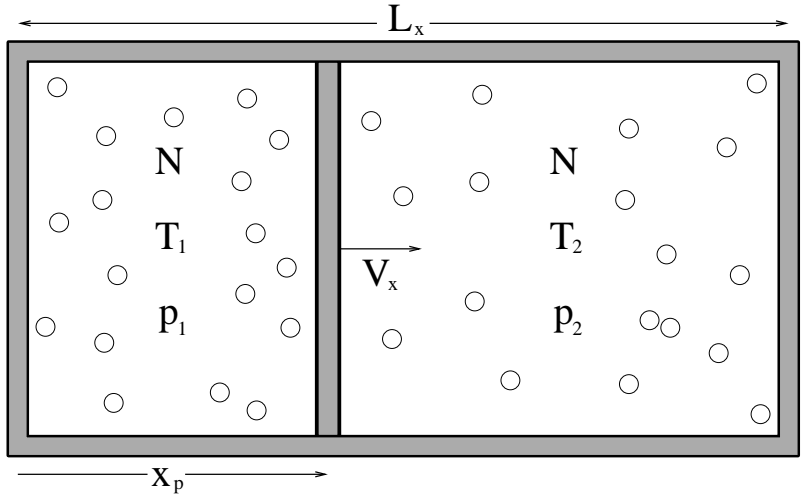}
\caption{Sketch of the adiabatic piston considered  here. Each of the compartments contains $N$ inelastic hard
spheres or disks of mass $m$ and diameter $\sigma$. The piston can move without friction along the $x$ axis, as indicated in the figure. \label{fig1}}
\end{figure}

The system considered consists of a cylinder of length $L_{x}$ filled with $2N$ inelastic hard spheres ($d=3$)
or disks ($d=2$) of mass $m$ and diameter $\sigma$. The cylinder is divided into two compartments by a piston
of mass $M$ perpendicular to the symmetry axis of the container, taken as the $x$-axis (Fig. \ref{fig1}). The
piston can move without friction along the $x$ direction, its motion being entirely induced by the collisions
with the gas particles. There are the same number of particles, $N$,  at each side of the piston.

The system is isolated and evolves freely in time. Collisions between particles are characterized by a constant
(velocity-independent) coefficient of normal restitution $\alpha$,
defined in the interval $ 0 < \alpha \leq 1$. On the other hand, collisions of the particles with the piston and with
the walls of the container are elastic. Attention will focus here on states in which the position of the piston has a stationary value
$x_{P}$.  In these states, mechanical equilibrium requires that the hydrodynamic pressure $p$ be the same at both sides
of the piston. The variables corresponding to each of the compartments will be identified by a subscript $i=1,2$,
respectively. It will be assumed that the gas remains always very dilute in both compartments and that, as a lowest
order approximation, the granular gas is homogeneous inside each compartment and it can be described by the equations
for the homogeneous cooling state (HCS) \cite{ByP04,Ca90}. This implies that the relative flux of energy through the piston
is very low as compared with the characteristic relaxation time of the HCS. Also, it requires that the system is not
too large so that the HCS is stable \cite{GyZ93} and the shearing and clustering instabilities are avoided. Then, for
the states under consideration, it is $p_{1}(t)=p_{2}(t)$ or, equivalently,
\begin{equation}
\label{2.1}
n_{1}T_{1}(t) = n_{2} T_{2}(t),
\end{equation}
where $n$ and $T$ denote the number density and the granular temperature respectively. Note that while the temperature
and the pressure will be time dependent, the stationarity of $x_{P}$ implies the same property for the density, since
\begin{equation}
\label{2.2}
n_{1}= \frac{N}{S x_{P}}, \quad n_{2} = \frac{N}{S(L_{x}-x_{P})}.
\end{equation}
Here $S$ is the section (area for $d=3$ and length for $d=2$) of the piston and the container. Energy balance equations
for each of the compartments when the piston is at rest are given by
\begin{equation}
\label{2.3}
\frac{d}{2} N \frac{dT_{i}}{dt}  = - \frac{d}{2} N \zeta_{i} T_{i}+ Q_{i}S,
\end{equation}
$i=1,2$. The first term on the right hand side is the Haff's law describing the cooling of the HCS due to the inelasticity
of collisions \cite{Ha83}. The explicit expression of the cooling rate is \cite{GyS95,vNyE98}
\begin{equation}
\label{2.4}
\zeta_{i}=\frac{p_{i}}{\eta_{0} (T_{i})} \zeta^{*} (\alpha),
\end{equation}
with $\eta_{0}$ being the elastic ($\alpha=1$) shear viscosity,
\begin{equation}
\label{2.5}
\eta_{0}(T)= \frac{d+2}{8} \Gamma \left( d/2 \right) \pi^{-(d+1)/2} (mT)^{1/2} \sigma^{-(d-1)},
\end{equation}
and $\zeta^{*}(\alpha)$ a dimensionless function of the restitution coefficient,
\begin{equation}
\label{2.6} \zeta^{\ast}(\alpha) =\frac{2+d}{4d}(1-\alpha^{2})
\left[1+\frac{3}{32} c^{*}(\alpha) \right]\, ,
\end{equation}
\begin{equation}
\label{2.7}
c^{\ast}(\alpha) \equiv
\frac{32(1-\alpha)(1-2\alpha^{2})}{9+24d+(8d-41)\alpha+
30\alpha^{2}(1-\alpha)}\, .
\end{equation}
The above expressions have been derived from the Boltzmann equation in the so-called
first Sonine approximation \cite{GyS95,vNyE98}.

The second term on the right hand side of Eq. (\ref{2.3}) is the energy flux going into the compartment $i$
through the piston velocity fluctuations. This quantity has been evaluated in \cite{ByR09} by means of a simplified kinetic theory in which both
the velocity distribution of the piston and of the inelastic gases were approximated by Gaussians. Particularization of Eq. (66) in
\cite{ByR09} for the present case of elastic collisions between the particles and the piston gives
\begin{equation}
\label{2.8}
Q_{i}= -2 \left( \frac{2}{\pi m} \right)^{1/2} \frac{M}{M+m} (1+ \phi_{i})^{1/2} \left[ 1- \frac{(1+ \phi_{i})M}{M+m}
\right] n_{i}T_{i}^{3/2},
\end{equation}
where
\begin{equation}
\label{2.9}
\phi_{i} \equiv \frac{m T_{P}}{MT_{i}}
\end{equation}
and $T_{P}$ is the temperature parameter of the piston defined from the second moment , $<V_{x}^{2}(t)>$, of
its velocity distribution by $M <V_{x}^{2}(t)> =T_{P}$. Of course, no macroscopic or thermodynamic interpretation
is assigned to this quantity, although it is expected to be a
decreasing function of time. More precisely, the energy balance equation for the piston yields
\begin{equation}
\label{2.10}
\frac{d T_{P}}{d t} = - 2 S (Q_{1}+ Q_{2}).
\end{equation}
Equations (\ref{2.1}), (\ref{2.3}), and (\ref{2.10}) form a closed set of four equations for the
unknown $x_{P}$, $T_{1}(t)$, $T_{2}(t)$, and $T_{P}(t)$.

To simplify the calculations and allow for an analytical solution, attention will be restricted in the following
to those cases where $m/M $ is small, so that, assuming that $T_{P}/T_{i}$ is of the order of unity (something to be checked
a posteriori), Eq. (\ref{2.8}) can be approximated by
\begin{equation}
\label{2.11}
Q_{i} \approx -2 \left( \frac{2m}{\pi} \right)^{1/2} \frac{T_{i}-T_{P}}{M} n_{i}T_{i}^{1/2}.
\end{equation}
The above expression is consistent with the intuitively expected behavior. Energy flux goes from
the system with the larger temperature parameter to the system with the lower one.

\section{The symmetric state}
\label{s3}
It is convenient to use a dimensionless time scale $s$ defined by
\begin{equation}
\label{3.1}
ds = (2n)^{1/2} \sigma^{d-1} \left[ \frac{p(t)}{m} \right]^{1/2} dt,
\end{equation}
where $n \equiv 2N/SL_{x}$ is the average number density of the whole system. Then, Eqs. (\ref{2.3}) and
(\ref{2.10}) read
\begin{equation}
\label{3.2}
\frac{d T_{i}}{ds}=- a_{i} T_{i} - \frac{b_{i}}{2Nd} (T_{i}-T_{P}),
\end{equation}$i=1,2$, and
\begin{equation}
\label{3.3}
\frac{dT_{P}}{ds} = \frac{b_{1}}{2} (T_{1}-T_{P}) + \frac{b_{2}}{2} (T_{2}-T_{P}).
\end{equation}
Here,
\begin{equation}
\label{3.4}
a_{i} \equiv \left( \frac{n_{i}}{n} \right)^{1/2} a, \quad
b_{i} \equiv \left( \frac{n_{i}}{n} \right)^{1/2} b,
\end{equation}
with $a$ and $b$ being constant parameters given by
\begin{equation}
\label{3.5}
a \equiv \frac{8 \pi^{(d+1)/2} \zeta^{*}}{(2+d) 2^{1/2} \Gamma (d/2)}\, ,
\end{equation}
\begin{equation}
\label{3.6}
b \equiv \frac{ 8 S m}{\pi^{1/2} \sigma^{d-1} M}\, .
\end{equation}
The temperatures $T_{1}$ and $T_{2}$ are not independent since they are related by Eq.\
(\ref{2.1}). Therefore, the three equations (\ref{3.2})-(\ref{3.4}) can be replaced by
the set of linear equations
\begin{equation}
\label{3.7}
\frac{dT_{1}}{ds}=- \left( a_{1}+\frac{b_{1}}{2Nd} \right) T_{1} + \frac{b_{1}}{2Nd}\, T_{P},
\end{equation}
\begin{equation}
\label{3.8}
\rho^{2} \frac{d T_{1}}{ds} = - \left( a_{1}+\frac{b_{1}}{2Nd} \right) \rho T_{1} + \frac{b_{1}}{2N \rho d}\, T_{P},
\end{equation}
\begin{equation}
\label{3.9}
\frac{dT_{P}}{ds} = \frac{b_{1}}{2} \left(  1 +\rho \right) T_{1} - \frac{b_{1}(1+ \rho)}{2 \rho}\, T_{P},
\end{equation}
where the relative density of the two compartments,
\begin{equation}
\label{3.10}
\rho^{2} \equiv \frac{n_{1}}{n_{2}}\, ,
\end{equation}
has been introduced. To search for solutions of the above equations write
\begin{equation}
\label{3.11}
T_{1}(s) = \theta_{1} e^{\lambda s}, \quad T_{P}(s) = \theta_{P} e^{\lambda s}.
\end{equation}
Then it follows that for the existence of solutions different from the trivial one
$\theta_{1}=\theta_{P}=0$ it must be
\begin{equation}
\label{3.12}
a_{1}+ \frac{b_{1}}{2Nd} + \lambda = \left( a_{1}+ \frac{b_{1}}{2Nd} + \lambda \rho \right) \rho^{2}
\end{equation}
and
\begin{equation}
\label{3.13}
\frac{2 \left( a_{1}+ \frac{b_{1}}{2Nd} + \lambda \right)}{b_{1} (1+ \rho)} = \frac{{b}_{1}}{2Nd \left[
\frac{b_{1} (1+\rho)}{2 \rho} + \lambda \right]}.
\end{equation}
Consider first the special case $\rho=1$, i.e. the symmetric state with $n_{1}=n_{2}=n$ and $x_{P}=
L_{x}/2$. Because of Eq.\ (\ref{2.1}) it is also $T_{1}(t)=T_{2}(t) = T(t)$ and, because of Eq.\
(\ref{2.11}), $Q_{1}=Q_{2}=Q$. For this value of $\rho$, Eq.\ (\ref{3.12}) becomes an identity, and only
the condition given by Eq.\ (\ref{3.13}) remains. The latter takes the form
\begin{equation}
\label{3.14}
\lambda^{2} + \left( a +b+ \frac{b}{2Nd} \right) \lambda + a b =0,
\end{equation}
and it has the solutions
\begin{equation}
\label{3.15}
\lambda_{\pm} = - \frac{c}{2} \pm \frac{ (c^{2}-4ab )^{1/2}}{2},
\end{equation}
\begin{equation}
\label{3.16}
c \equiv a + b \left( 1 + \frac{1}{2Nd} \right).
\end{equation}
It is easy to verify that $\lambda_{-} < \lambda_{+} <0$ and, therefore, for large enough times
the temperatures of the granular gases and the piston in the symmetric state have the form
\begin{equation}
\label{3.17}
T^{(s)}(s)= \theta^{(s)} e^{\lambda^{(s)} s}, \quad T_{P}^{(s)}(s)= \theta_{P}^{(s)} e^{\lambda^{(s)} s},
\end{equation}
respectively, with $\lambda^{(s)}= \lambda_{+}$. The coefficients $\theta^{(s)}$ and $\theta_{P}^{(s)}$
are related through
\begin{equation}
\label{3.18}
\theta^{(s)}= \left( 1 + \frac{\lambda^{(s)}}{b} \right) \theta_{P}^{(s)},
\end{equation}
that follows from Eq.\ (\ref{3.9}) after particularizing it for the symmetric state. Using Eq.\
(\ref{3.15}) it can be checked that $|\lambda^{(s)}|<b$, as required by consistency, since otherwise
the theory would predict negative values for one of the temperature parameters. In addition, Eqs.\
(\ref{3.17}) and (\ref{3.18}) imply that
\begin{equation}
\label{3.19}
T_{P}^{(s)}(t) > T^{(s)} (t),
\end{equation}
i.e. the temperature of the granular gases is smaller than the temperature parameter of the piston at the
same time. This reflects that the physical mechanisms for which the piston cools is the energy dissipation in
the gas collisions, so that $T_{P}^{(s)}$ is driven by $T^{(s)}$.

Equations (\ref{3.17}) indicate that the temperature parameters of both the piston and the granular gases cool
at the same rate,
\begin{equation}
\label{3.20}
\frac{dT^{(s)}}{dt} = - \zeta_{ef}^{(s)} T{(s)}, \quad \frac{dT_{P}^{(s)}}{dt} = - \zeta_{ef}^{(s)} T^{(s)},
\end{equation}
where
\begin{equation}
\label{3.21}
\zeta_{ef}^{(s)} = - \lambda^{(s)} \left( \frac{2T}{m} \right)^{1/2} n \sigma^{d-1}
\end{equation}
is the effective cooling rate of the gas taking into account the energy interchange with the piston. In the
limit $N \rightarrow \infty$ with all the other parameter remaining finite, it is $\lambda^{(s)} \approx -a$ and $\zeta_{ef}^{(s)} \approx \zeta$, as expected.
In the same limit, Eq.\ (\ref{3.18}) leads to
\begin{equation}
\label{3.22}
\theta^{(s)} \approx \left( 1 - \frac{a}{b} \right) \theta_{P},
\end{equation}
showing that the symmetric state does not exist if $a>b$ in this limit. But for $M/m \gg 1$, it is $b \ll 1$. This means that
for $N \rightarrow \infty$, $M/m \rightarrow \infty$, the state with $x_{P} = L_{x} /2$ does not exist. The physical reason is that when the piston has a very large mass and the number of particles is very large, the piston is not able to cool down with the
same rate as the granular gas.

\section{The asymmetric state}
\label{s4}
To investigate the existence of states with a stationary position of the piston different from
the middle of the system, Eqs.\ (\ref{3.12}) and (\ref{3.13}) have to be solved. For $\rho \neq 1$, eq.
(\ref{3.12}) yields
\begin{equation}
\label{4.1}
\lambda = \lambda^{(a)} = \left(a + \frac{b}{2Nd} \right) \left( \frac{n_{1}n_{2}}{n} \right )^{1/2}
\frac{n_{2}-n_{1}}{n_{1}^{3/2}-n_{2}^{3/2}}.
\end{equation}
In the above expression $\rho$ has been eliminated in favor of the number densities to emphasize the invariance
with respect to the interchange $n_{1} \leftrightarrow n_{2}$, as required by symmetry. Therefore, it is
\begin{equation}
\label{4.2}
T_{1}^{(a)}(s) =\theta_{1}^{(s)} e^{\lambda^{(a)} s}, \quad
T_{2}^{(a)}(s) =\theta_{2}^{(s)} e^{\lambda^{(a)} s},
\end{equation}
\begin{equation}
\label{4.3}
T_{P}^{(s)} = \theta_{P}^{(s)} e^{\lambda^{(a)} s},
\end{equation}
with
\begin{equation}
\label{4.4}
\frac{\theta_{1}^{(a)}}{\theta_{2}^{(a)}} = \frac{1}{\rho^{2}}
\end{equation}
and
\begin{equation}
\label{4.5}
\frac{\theta_{P}^{(a)}}{\theta_{1}^{(a)}}= \left( 1 + \frac{2Nad}{b} \right) \frac{\rho^{2}}{\rho^{2}+ \rho +1}\, .
\end{equation}

The last relation follows directly from Eq.\ (\ref{3.7}). Still, it must be required that Eq.\ (\ref{3.13}) be fulfilled but, before doing that, it is important to realize that Eqs. (\ref{4.2}) and (\ref{4.3}) lead to the result that the cooling rate of both granular gases and also the piston are again the same in the asymmetric state(s), namely
\begin{equation}
\label{4.6}
\frac{dT_{i}^{(a)}}{dt}= - \zeta_{ef}^{(a)} T_{i}^{(a)}(t),
\end{equation}
\begin{equation}
\label{4.7}
\frac{dT_{P}^{(a)}}{dt}= - \zeta_{ef}^{(a)} T_{P}^{(a)}(t),
\end{equation}
\begin{equation}
\label{4.8}
\zeta_{ef}^{(a)} = - \lambda^{(a)} \left( \frac{2 p(t) n}{m} \right)^{1/2}  \sigma^{d-1}.
\end{equation}

Substitution of Eq. (\ref{4.1}) into Eq.\ (\ref{3.13}), after some algebra leads to
\begin{equation}
\label{4.9}
z^{2}-N d z +Nd=0,
\end{equation}
with
\begin{equation}
\label{4.10}
z \equiv  \left( \frac{2Nad}{b}+1 \right) \frac{ \rho}{\rho^{2}+\rho +1}\, .
\end{equation}
For large values of $N$, the two solutions of Eq. (\ref{4.9}) read
\begin{equation}
\label{4.11}
z_{1} \approx 1, \quad z_{2} \approx Nd.
\end{equation}
The next task is to invert Eq. (\ref{4.10}) to get the values of the density ratio $\rho$.
Consider first the root $z_{1}=1$. Because of Eq. (\ref{4.5}) it is
\begin{equation}
\label{4.12}
\theta_{P}^{(a)}= \rho \theta_{1}^{(a)}= \left( \theta_{1}^{(a)} \theta_{2}^{(a)} \right)^{1/2},
\end{equation}
or, in terms of the actual temperature parameters,
\begin{equation}
\label{4.13}
T_{P}^{(a)}(t) = \left[ T_{1}^{(a)}(t) T_{2}^{(a)}(t) \right]^{1/2}.
\end{equation}
For $z=1$, Eq.\ (\ref{4.10}) becomes
\begin{equation}
\label{4.14}
\rho^{(a)2}- \frac{2Nad}{b}\, \rho^{(a)} +1 =0,
\end{equation}
whose solution is
\begin{equation}
\label{4.15}
\rho^{(a)}= \frac{Nad}{b} + \left[ \left( \frac{Nad}{b} \right)^{2}-1 \right]^{1/2}.
\end{equation}
(The other solution of the second degree equation is the inverse of the above and corresponds
to the interchange of the compartments $1$ and $2$). Existence of a physical solution requires that
\begin{equation}
\label{4.16}
\frac{Nad}{b} >1.
\end{equation}
Using Eqs.\ (\ref{3.5}) and (\ref{3.6}), this condition can be expressed as
\begin{equation}
\label{4.17}
M>M_{c},
\end{equation}
where
\begin{equation}
\label{4.18}
M_{c} \equiv \frac{(d+2) 2^{1/2} \Gamma \left( d/2 \right) Sm}{
d \pi^{d/2} \sigma^{d-1} N \zeta^{*}}
\end{equation}
can be understood as a critical value of the mass of the piston below which the asymmetric state with $z=1$
is not possible. Moreover, the larger $M/M_{c}$ the stronger the asymmetry of the state as measured by the
position of the piston $x_{P}$. In the limit $ M/M_{c} \rightarrow \infty$ , either $x_{P} \rightarrow 0$ or
$x_{P} \rightarrow L_{x}$. This is just the limiting situation discussed in ref. \cite{BRyvB05}.

It is relevant to  see what happens at the bifurcation point, i.e. for $M \rightarrow M_{c}$ or, more precisely,
for $M-M_{c} \rightarrow 0^{+}$. In this limit $\rho \rightarrow 1$, and Eqs. (\ref{4.2})-(\ref{4.5}) yield
\begin{equation}
\label{4.19}
T_{1}^{(a)}(t) = T_{2}^{(a)}(t)=T_{P}^{(a)}(t).
\end{equation}
Moreover, Eq.\ (\ref{4.1}) gives $\lambda^{(a)} \rightarrow -a$. This value agrees with the relaxation
rate $\lambda^{(s)}$ of the symmetric state, given in Eq. (\ref{3.15}) in the limit $Nd \gg 1$, indicating the continuity of the cooling rate through the transition.

Still remains to be studied the solution $z_{2} \approx Nd$ of Eq.\ (\ref{4.9}). It is easily seen that this value leads to the relation
\begin{equation}
\label{4.20}
T_{P}(t)^{(a) \prime} = Nd \left[ T_{1}^{(a)\prime}(t)T_{2}^{(a)\prime}(t) \right]^{1/2}\, ,
\end{equation}
following from Eq.\ (\ref{4.5}). It follows that this solution has to be discarded from our analysis since for
large $N$ the temperature parameter of the piston is much larger that the temperature of the granular gases to
both sides of it, something that contradicts the assumption made to derive the approximate expression  for
$Q_{i}$ given in Eq.\ (\ref{2.11}) and used throughout the paper.

\section{Molecular Dynamics simulation results}
\label{s5}

To check the above theoretical predictions, molecular dynamics (MD) simulations with an event driven algorithm \cite{AyT87,PyS05} have been performed. The system considered is an ensemble of $2N$ inelastic disks enclosed in a two-dimensional rectangular box of sides $L_{x}$ and $L_{y}$. The latter corresponds to the general transversal section $S$ used in the previous sections. Collisions of the particles with the walls of the container and also with the piston are elastic. In all the simulations, the piston was initially placed in the middle of the container and the particles were evenly and uniformly distributed between the two compartments. The initial number density was always quite low. Moreover, the initial velocity distributions of the disks  were Gaussian and with the same temperature at both sides of the piston. On the other hand, the latter was placed with a vanishing initial velocity.

There is a limitation in the systems that can be simulated as a consequence of the instability of the HCS. As the value of the coefficient of normal restitution $\alpha$ decreases, the critical size for which the HCS becomes unstable also decreases \cite{GyZ93,BRyM98}, that means that for a given density the number of particles becomes smaller. In the low density limit, this leads very soon to a very reduced number of particles, so that boundary effects becomes very relevant, specially when dealing with a system having walls (as opposite to periodic boundary conditions), as it happens in the present case. As a consequence, it turns out that the particular states addressed here are not reached for values of $\alpha$ roughly smaller than $0.85$.

\begin{figure}
\includegraphics[scale=0.7,angle=0]{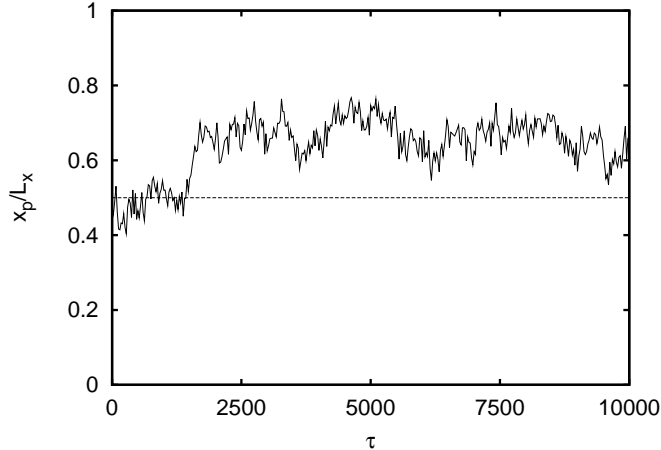}
\caption{Time evolution of the position of the piston for a particular simulation trajectory for a system
with $2N =200 $, $\alpha= 0.98$, $L_{x} =  2L_{y} =  100\sigma$, and $M=25 m$. Time $\tau$ is measured in accumulated number of collisions per particle. The dashed line is a guide for the eye. \label{fig2}}
\end{figure}

In some cases, it was observed that the piston stays oscillating around the initial position, $x_{P}= L_{x}/2$. These oscillations are easily understood as being induced by the pressure fluctuations of the granular gas. This behavior persists in time, so that the initial symmetry of the system is conserved. It is clear that these situations correspond to the symmetric state discussed in Sec.\ \ref{s3}. On the other hand, in other cases the initial oscillations increase in time and after a while they  combine with a net motion of the piston moving away from the middle of the system. Eventually, the net motion ceases and a steady average position is observed. The system reaches one of the asymmetric states analyzed in Sec. \ref{s4}. Figure \ref{fig2} shows a typical example of the observed behavior when the initial symmetry is broken. The parameter values in this case are $\alpha = 0.98$, $2N=200 $, $L_{x} = 2 L_{y} = 100 \sigma$, and  $M =25 m$. Then, the initial number density is $n_{1}=n_{2}= 0.04\sigma^{-2}$. Note that the time scale $\tau$ used in the figure is the accumulated number of collisions per particle. In the particular simulation reported, the piston
moves towards the right compartment, reaching a stationary average position at $x_{P} \approx 0.66 L_{x}$. In different simulation realizations with the same values of the parameters, the final position of the piston occurs equally often
to each side of the middle, but always at the same average distance of it. Moreover, it is worth to mention that
in some realizations spontaneous transitions between the two average positions at both sides of the middle were observed.

\begin{figure}
\includegraphics[scale=0.7,angle=0]{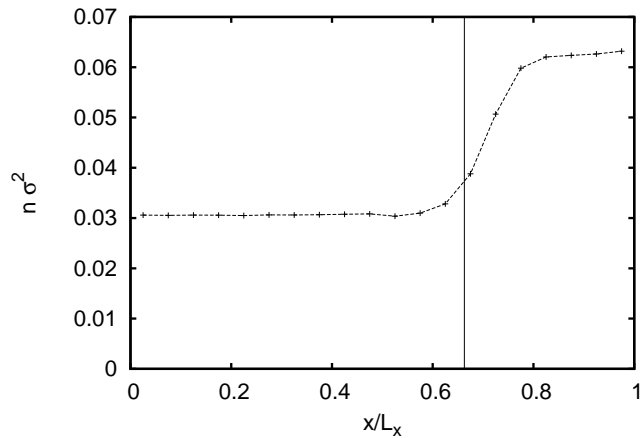}
\caption{Steady density profile exhibited by the  the system with the same parameters as in
Fig. \protect{\ref{fig2}. The solid vertical line indicates the steady average position of the piston.  \label{fig3}}}
\end{figure}

When the average position of the piston reaches a steady value, also the average density profile becomes time independent. It is given in Fig. \ref{fig3} for the same values of the parameter of the system as in Fig. \ref{fig2}. The plotted curve is an average on time once in the steady configuration and also over
$350$ simulation trajectories. It is observed that outside a boundary layer at both sides of the
piston, the density can be considered as uniform and different in each of the two compartments, in agreement with the assumption made
in the theoretical model developed in Sec. \ref{s2}.  The density boundary layer is mainly due to the oscillations of the piston.

\begin{figure}
\includegraphics[scale=0.7,angle=0]{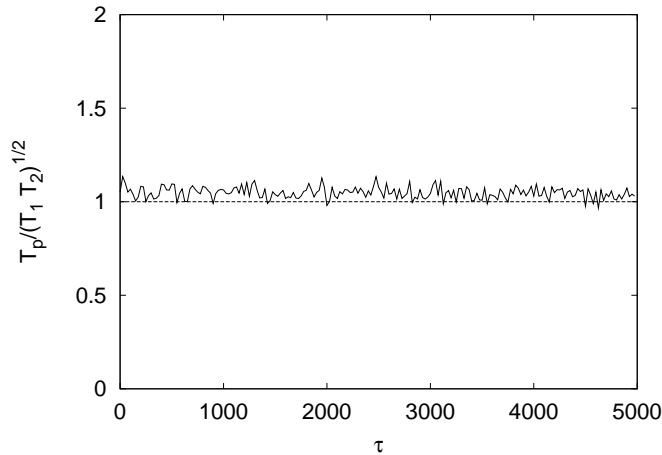}
\caption{Temperature of the piston $T_{p}$ divided by the square root of the product of the temperatures of the
granular gases in each of the compartments, $T_{1}$ and $T_{2}$, respectively, for the same system as in Figs. \protect{\ref{fig2}}
and \protect{\ref{fig3}}. The data have been averaged over $2000$ trajectories. Time is measured in accumulated number
of collisions per particles.  \label{fig4}}
\end{figure}

Another strong prediction of the theory is Eq.\ (\ref{4.13}), relating the steady temperature of the piston and those of
the inelastic gases in both compartments. The MD simulations show that the relation is verified in the range of parameters in which the asymmetric state has been observed. In Fig. \ref{fig4}, this is illustrated for the same values of the parameters as in Figs. \ref{fig2} and \ref{fig3}. To put the comparison in a proper context, it must be taken into account that  along every trajectory the piston jumps several times from the average position at one side of the system to the symmetric one, through states which are not described by the theory developed here. These configurations have not be removed when computing the results reported in the figure.  It is worth to notice that Eq.\ (\ref{4.13}) is consistent with the result that the cooling rates of the gases and the piston are the same.

\begin{figure}
\includegraphics[scale=0.7,angle=0]{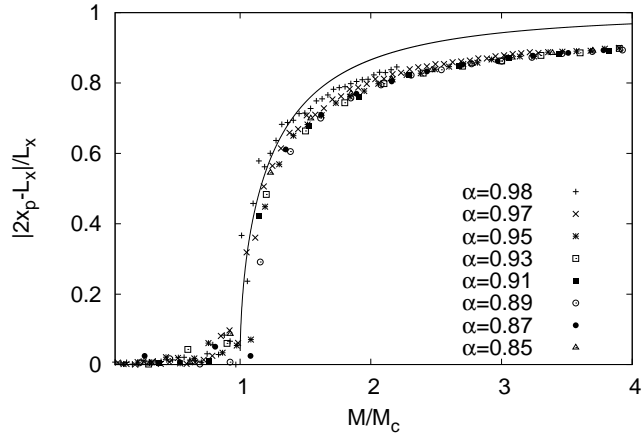}
\caption{Bifurcation diagram showing the asymmetry of the number density  as a function of the dimensionless control
parameter $M/M_{c}$, where $M$ is the mass of the piston and $M_{c}$ is given by Eq.\ (\protect{\ref{4.18}}). The symbols are from molecular dynamics simulations and the solid line is the theoretical prediction derived in the paper, Eqs.\ (\protect{\ref{5.2}}) and (\protect{\ref{5.3}}). All the  simulation results reported in this figure have been obtained with $L_{x}=2 L_{y}=100 \sigma$ and $2N=200$, so that $n \sigma^{2} =0.04$. Different values of $\alpha$ have been used as indicated in the insert. For each value, the mass ratio $M/m$ has been varied in order to change the value of $M/M_{c}$. \label{fig5}}
\end{figure}

\begin{figure}
\includegraphics[scale=0.7,angle=0]{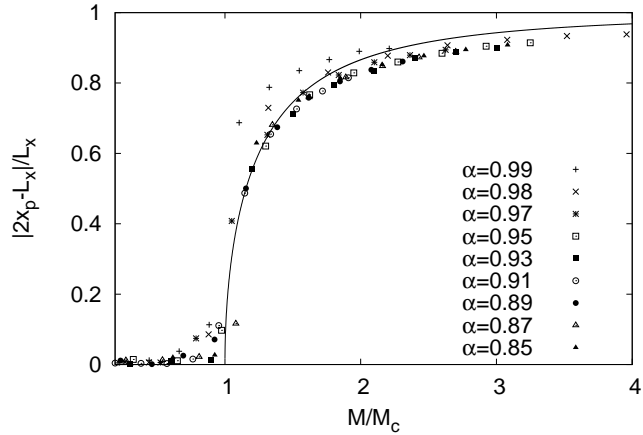}
\caption{The same as in Fig.\ \protect{\ref{fig5}}, with the only difference that in the simulations reported here it is $L_{x}=4L_{y} = 200 \sigma$ and, consequently, $  n \sigma^{2} =0.02$. \label{fig6}}
\end{figure}

Therefore, the simulations clearly show the existence of both the symmetric and the asymmetric states. In the latter, the steady average position of the piston can take any value inside the system, depending on the values of the
parameters. In order to carry out a more quantitative check of the theory, the average position of the piston $x_{P}$ has been measured as a function of the ratio $M/M_{c}$, where $M_{c}$ is the parameter defined in Eq.\ (\ref{4.18}). The results are plotted in Figs.\ \ref{fig5} and \ref{fig6} in terms of the asymmetry parameter
\begin{equation}
\label{5.1}
\epsilon \equiv \frac{ |2x_{P}-L_{x}|}{L_{x}} =\frac{|1-\rho^{2}|}{1+\rho^{2}}\,
\end{equation}
defined in the interval $0 \leq \epsilon \leq 1$. The prediction for this quantity obtained here is
\begin{equation}
\label{5.2}
\rho = \rho^{(s)}= 1 \textrm{ and } \epsilon=0 \textrm{ for } M<M_{c}
\end{equation}
\begin{equation}
\label{5.3}
\rho = \rho^{(a)} = \frac{M}{M_{c}}+ \left[ \left( \frac{M}{M_{c}} \right)^{2} -1 \right]^{1/2} \textrm{ for } M>M_{c}.
\end{equation}
The observed agreement between theory and simulations can be considered satisfactory. In particular, the bifurcation
point seems to be quite accurately predicted by the theory. Upon evaluating the results, it must be taken into account that when the asymmetry of the state increases the density in one of the compartments also increases, and that the
theory developed here is based on the assumption that the granular gas in both compartments can be treated in the very dilute limit approximation.

\section{Concluding remarks}
\label{s6}
The spontaneous symmetry breaking investigated here has some similarities with  phenomena of phase
separation occurring in vibrofluidized granular materials, both in presence \cite{Eg99,vWvMVyL01} and absence \cite{BMGyR01} of gravity. There are also significant differences. A relevant one is that the
system studied in this paper is isolated and has no stationary states, as a consequence of the energy dissipation
in collisions. The states addressed are characterized by a stationary position of the movable piston dividing the system into two parts. A main result obtained is that the conditions determining these states do not require
the (granular) temperature to be the same at both sides of the piston, as it is the case in normal, elastic fluids.
Instead, the characteristic relaxation times of the energy must be the same in both compartments, and agree with the one of the piston. These times can be interpreted as the
effective cooling rates of the granular gases. Moreover, the gases tend to be homogenous at both sides of the piston, although presenting boundary layers next to it.

There is an interesting physical picture emerging from the above results. It is well known that when two molecular systems are put into contact through a movable thermal wall, keeping the whole ensemble isolated, the final state is of equilibrium, implying that each of the two macroscopic subsystems is also at equilibrium, having both the same pressure and temperature.
If the same experiment is carried out with granular gases, the results being reported  indicate that the whole system tends to a
cooling state characterized by a unique time-dependent  temperature parameter $T(t)$ and being homogeneous inside each compartments. To be more precise, what is meant is that at a microscopic level the probability distribution function of the system has the form
\begin{equation}
\label{6.1}
f(\Gamma,t)=\left[ v(t) \right]^{-(2Nd+1)} f^{*} \left( \left\{ {\bf r}_{i} \right\}, X, \left\{ \frac{{\bm v}_{i}}{v(t)} \right\}, \frac{V_{x}}{v(t)} \right),
\end{equation}
where $\Gamma$ denotes a point of the phase space of the system, ${\bm r}_{i}$ and ${\bm v}_{i}$ ($i=1, \ldots, 2N$) are the position and velocity of particle $i$, $X$ and $V_{x}
$ are the position and velocity of the piston, and $v(t) \equiv \left( 2T(t)/m \right)^{1/2}$. The peculiarity of this state is that all its time dependence occurs through the parameter $T(t)$. The function $f^{*}$ includes the constraint  that some given  particles are located at one side of the piston and the remaining particles at the other one. Temperature parameters $T_{i}(t)$, $i=1,2$, for each compartment can be defined in the usual way from the second
velocity moment of its particles. Also a temperature parameter $T_{P}(t)$ for the piston can be defined in a similar way. The scaling form of Eq.\ (\ref{6.1}) implies that
\begin{equation}
\label{6.2}
\frac{d \ln T_{1}(t)}{dt}= \frac{d  \ln T_{2}(t)}{dt}= \frac{d  \ln T_{P}(t)}{dt},
\end{equation}
i.e., the cooling rates of both compartments and also of the piston are the same. This is the result found in this paper. Trivially, Eq. (\ref{6.2}) is verified if all the temperatures are the same, i.e. $T_{1}(t)=T_{2}(t)=T_{P}(t)$.
This is what happens in equilibrium molecular systems, where velocity correlations between the three macroscopic systems are negligible and the equilibrium distribution for the velocity factorizes. On the other hand, when dealing with granular gases, velocity correlations between the two compartments and the piston are relevant, the velocity distribution of the whole system does not factorize and the three temperature parameters differ.

The above feature is related with the non-equipartition of kinetic energy in mixtures of granular gases. The granular temperature of the components of a mixture defined from the average kinetic energy of each species is not the same
\cite{JyM87,GyD99}. The homogeneous cooling state of the mixture has the property that the cooling rates for all the partial temperatures are the same. This condition determines all the temperatures of the components in terms of a unique temperature parameter. Actually, a similar result holds in vibrated granular systems, in the sense  that only a temperature is needed for a macroscopic description of mixtures of granular gases \cite{ByR09a}. Here, the result is extended to spatially separated granular gases  interacting through a movable piston.

A consequence of the cooling rate being the same as compared with the temperatures being the same, is the existence of the spontaneous symmetry breaking discussed in this paper. The phenomenon can be easily understood in terms of
rather simple kinetic theory arguments, whose theoretical predictions are in good agreement with molecular dynamics simulation results. An interesting issue to be addressed in the near future is why the asymmetric state is more stable than the symmetric one when both states exist. Perhaps this is related with some extremal principle that shed light on the macroscopic physics of granular fluids.

\section{Acknowledgements}

This research was partially supported by the Ministerio de
Educaci\'{o}n y Ciencia (Spain) through Grant No. FIS2008-01339 (partially financed
by FEDER funds).

\end{document}